\documentclass[conference]{IEEEtran}
\IEEEoverridecommandlockouts
\usepackage{cite}
\usepackage{amsmath,amssymb,amsfonts}
\usepackage{algorithmic}
\usepackage{graphicx}
\usepackage{textcomp}
\usepackage{listings}
\usepackage{xcolor}
\usepackage{listings}
\usepackage[hidelinks]{hyperref}
\usepackage{comment}

\usepackage{todonotes}
\newcommand{\gbnote}{\todo[inline]} %comments from Gabriella
\usepackage{url}
\def\BibTeX{{\rm B\kern-.05em{\sc i\kern-.025em b}\kern-.08em
    T\kern-.1667em\lower.7ex\hbox{E}\kern-.125emX}}

\usepackage{xcolor}
\definecolor{listinggray}{gray}{0.9}
\definecolor{lbcolor}{rgb}{0.9,0.9,0.9}
\usepackage{listings}
\usepackage{amsfonts}

\definecolor{codegreen}{rgb}{0,0.6,0}
\definecolor{codegray}{rgb}{0.5,0.5,0.5}
\definecolor{codepurple}{rgb}{0.58,0,0.82}
\definecolor{backcolour}{rgb}{0.95,0.95,0.92}
 \usepackage{multicol}
\lstdefinestyle{mystyle}{
    backgroundcolor=\color{backcolour},  
    commentstyle=\color{codegreen},
    keywordstyle=\color{magenta},
    numberstyle=\tiny\color{codegray},
    stringstyle=\color{codepurple},
    basicstyle=\footnotesize,
    breakatwhitespace=false,        
    breaklines=true,                
    captionpos=b,                   
    keepspaces=true,                
    numbers=left,                   
    numbersep=5pt,                 
    showspaces=false,               
    showstringspaces=false,
    showtabs=false,                 
    tabsize=2
}
 
\begin{document}
\bstctlcite{IEEEexample:BSTcontrol}

\title{Dynamic Solutions for Hybrid Quantum-HPC Resource Allocation

% Commented Thanks block as we are in a Double Blind submission

\thanks{Funded by the European Union – NextGenerationEU: ICSC National Centre, CN00000013, MUR Act n. 1031 - 17/06/2022; and Barcelona Zettascale Lab, promoted by the Ministry for Digital Transformation and the Civil Service, within the framework of the Recovery, Transformation, and Resilience Plan.
BSC researchers are also supported by the Dpt. of Research and Universities of the Government of Catalonia to the AccMem (Code: 2021 SGR 00807).
Antonio J. Peña was partially supported by the Ramón y Cajal fellowship RYC2020-030054-I funded by MCIN/AEI/ 10.13039/501100011033 and by ``ESF Investing in Your Future''.
}
}

%Commenting authors as we are in a double blind submission
\author{\IEEEauthorblockN{Anonymous Authors}}

\author{\IEEEauthorblockN{
Roberto Rocco\IEEEauthorrefmark{1}\IEEEauthorrefmark{4},
Simone Rizzo\IEEEauthorrefmark{1},
Matteo Barbieri\IEEEauthorrefmark{1},
Gabriella Bettonte\IEEEauthorrefmark{1},
Elisabetta Boella\IEEEauthorrefmark{1},\\
Fulvio Ganz \IEEEauthorrefmark{1},
Sergio Iserte\IEEEauthorrefmark{8},
Antonio J. Peña\IEEEauthorrefmark{8}, 
Petter Sandås\IEEEauthorrefmark{8}, 
Alberto Scionti\IEEEauthorrefmark{2},
Olivier Terzo\IEEEauthorrefmark{2},\\
Chiara Vercellino\IEEEauthorrefmark{2}\IEEEauthorrefmark{5}, 
Giacomo Vitali\IEEEauthorrefmark{2}\IEEEauthorrefmark{5},
Paolo Viviani\IEEEauthorrefmark{2},
Jonathan Frassineti \IEEEauthorrefmark{3},
Sara Marzella\IEEEauthorrefmark{3},\\
Daniele Ottaviani\IEEEauthorrefmark{3},
Iacopo Colonnelli\IEEEauthorrefmark{7},
Daniele Gregori\IEEEauthorrefmark{1}
}\\
\IEEEauthorrefmark{1}\textit{E4 Computer Engineering}, Scandiano, Italy
\IEEEauthorrefmark{2}\textit{LINKS Foundation}, Torino, Italy\\
\IEEEauthorrefmark{8}\textit{Barcelona Supercomputing Center (BSC-CNS)}, Barcelona, Spain
\IEEEauthorrefmark{3}\textit{CINECA}, Casalecchio di Reno, Italy\\
\IEEEauthorrefmark{5}\textit{Politecnico di Torino}, Torino, Italy
\IEEEauthorrefmark{7}\textit{Università di Torino}, Torino, Italy\\

\IEEEauthorrefmark{4}\textit{roberto.rocco@e4company.com}\\\vspace*{-1cm}
}

\maketitle

\begin{abstract}

The integration of quantum computers within classical High-Performance Computing (HPC) infrastructures is receiving increasing attention, with the former expected to serve as accelerators for specific computational tasks. However, combining HPC and quantum computers presents significant technical challenges, including resource allocation. This paper presents a novel malleability-based approach, alongside a workflow-based strategy, to optimize resource utilization in hybrid HPC-quantum workloads. With both these approaches, we can release classical resources when computations are offloaded to the quantum computer and reallocate them once quantum processing is complete. Our experiments with a hybrid HPC-quantum use case show the benefits of dynamic allocation, highlighting the potential of those solutions.

\end{abstract}

\IEEEpeerreviewmaketitle

\begin{IEEEkeywords}
HPC, Quantum Computing, Dynamic Resource Management, Malleability
\end{IEEEkeywords}

\section{Introduction}
%\textcolor{blue}{We should check the verbs. In the text sometimes we use present, sometimes past, sometimes present perfect - we should uniform this.}

Quantum Computing (QC) technologies have evolved significantly, making quantum utility look attainable soon. As performance improves, the research community has begun to explore how quantum computers might fit into the broader computing continuum. Quantum systems are particularly well suited for addressing specific classes of NP-hard problems~\cite{Au_Yeung_2023}. This feature makes them ideal candidates to complement traditional High-Performance Computing (HPC) systems. In this emerging paradigm, it is widely anticipated that quantum processing units (QPUs) will serve as accelerators for computationally intensive, exponentially scaling workloads within scientific applications~\cite{ruefenachtBringingQuantumAcceleration,beck2024quantum,Elsharkawy_2024}, while HPC systems will provide the necessary scalability to support hybrid quantum-classical applications~\cite{mantha_2024}. 

Bringing QC into HPC environments holds promise for boosting progress across many scientific domains. Beck~\textit{et~al.} have recently identified a set of applications spanning various scientific fields, which would highly benefit from an HPC-QC integration~\cite{beck2024quantum}. These applications include quantum many-body dynamics for simulating quantum systems, continuum mechanics simulations using quantum linear solvers for problems like fluid dynamics, quantum-enhanced machine learning for developing advanced models, and quantum optimization to tackle complex optimization problems. Early examples of hybrid HPC-QC algorithms have begun to appear in the literature. Vercellino~\textit{et~al.} developed a graph coloring code relevant in several industrial contexts by combining a parallel branch-and-bound algorithm with a quantum routine that solves the maximum independent set problem to sample potential coloring solutions~\cite{Vercellino_2023}. Similarly, Kim and Suh considerably enhanced the performance of their hybrid optimization algorithm for metamaterial design by parallelizing via Message Passing Interface (MPI) both the machine learning phase preceding and the wave-optics simulation phase following the quantum approximate optimization algorithm~\cite{kim_2024}.

%Quantum computing technologies have evolved significantly recently, approaching a point where quantum utility appears attainable in a few years. Alongside performance improvements, the community is starting to envision the future role of quantum nodes in the computing continuum: right now, quantum computers are a standalone resource accessed through the cloud, but that is not sufficient for the future, as their computations will produce a large quantity of data that would be too burdensome to transfer over the network. For this reason, it becomes natural to integrate quantum resources with other computation resources that can handle the vastity of data produced promptly, i.e., HPC nodes. When integrated with HPC resources, QC acts as an accelerator, providing their performance on compatible algorithms while leaving the communication and coordination to classical resources. 

For all these reasons, a heterogeneous HPC-QC cluster is an appealing prospect. However, realizing such integration presents significant challenges due to the disparity in technological maturity between HPC and QC, their fundamentally different architectural paradigms, and the substantial imbalance in resource availability between classical and quantum systems~\cite{Elsharkawy_2023}. Among these challenges, efficient resource management and allocation stand out as critical concerns~\cite{Elsharkawy_2023, mantha_2024, shehata_2025, döbler2025surveyintegratingquantumcomputers}. In particular, dedicated scheduling strategies are essential to prevent resource under-utilization, especially in scenarios where only one or two QPUs are available per cluster~\cite{schulz_2022, brittHighPerformanceComputingQuantum2017,saurabh_2023}. A recent study analyzed and discussed potential solutions for efficient resource allocation in HPC-QC integration, proposing malleability to minimize computational resource waste~\cite{Simone}. A malleable job is capable of dynamically adjusting its resource allocation at runtime. In the HPC-QC context, this would enable a job to release classical HPC resources during quantum computation phases and reallocate them once computation returns to the classical phase.

In this work, we explore for the first time the use of \emph{malleability} in hybrid HPC-QC jobs to ensure efficient resource utilization. We also examine the potential of achieving effective resource management through a \emph{workflow}-based approach. It is worth noting that, although the use of Workflow Management Systems (WMSs) has been proposed for some time~\cite{osti_2006942}, there are still few examples of hybrid HPC-QC workloads that leverage WMSs to optimize resource usage in such heterogeneous environments~\cite{CRANGANORE2024346,DeMaio_2024}. As such, our use case implementation based on StreamFlow~\cite{Colonnelli:2021} introduces original elements and can help promote adopting this approach within the HPC-QC context.

To support this study, we parallelized an existing classical-quantum application to perform clustering aggregation of data leveraging different algorithms~\cite{Li2012ClusteringAA,scotti2024clusteringaggregationalgorithmneutralatoms}. The application provides the ideal use case to show the interaction between HPC and QC resources since the clustering methods are executed in parallel on HPC resources, and the aggregation occurs on the QC resource. We test our application on an \textit{ad-hoc} cluster comprising three classical compute nodes, with an additional classical node functioning as a quantum emulator. We then extend the application to support malleability by integrating the Dynamic Management of Resources (DMR) framework~\cite{iserte_dmrlib_2020}. Finally, we evaluate time-to-solution and resource usage across three scenarios: (1) no hybrid resource management; (2) hybrid resource management using a workflow-based approach implemented with the StreamFlow engine~\cite{Colonnelli:2021}; and (3) hybrid resource management with malleability.

Our experiments demonstrate that workflow-based and malleability-based approaches significantly reduce resource usage, effectively minimizing waste. Furthermore, we show that the malleability approach optimizes resource utilization in general and results in shorter time-to-solution than the workflow-based method. Notably, this temporal gain becomes increasingly pronounced as the computational load on the cluster grows.
%Moreover, we consider the resource allocation issues arising from the disparity in resource availability between classical and quantum resources, designing and implementing solutions to ensure proper resource usage. We test our solutions on an ad-hoc cluster featuring a quantum emulator, showing the benefits of our approaches and the potential of hybrid applications. 

Overall, the contributions of this paper are the following:
\begin{itemize}
\item We overview the design of hybrid HPC-QC applications starting from classical-quantum ones, evaluating different resource access schemes on a practical example;
\item We apply a novel solution based on malleability to address the issue of resource allocation in an HPC-QC environment with limited QC resources;
\item We compare our results across three test scenarios: without hybrid resource management, with workflow-based management, and with malleability.
\end{itemize}

This paper is structured as follows.
Section~\ref{sec:hybrid} illustrates, through a concrete example, the transformation of a classical-quantum application into an HPC-QC workload, highlighting the benefits of integrating HPC and QC in real-world scenarios; Section~\ref{sec:access} overviews the proposed approaches to deal with hybrid resource allocation;
Section~\ref{sec:exp} presents our experimental campaign, along with a discussion of the results. Finally, Section~\ref{sec:conclusion} concludes the article.

\section{Towards HPC-QC: clustering aggregation}\label{sec:hybrid}

In recent years, a growing number of hybrid applications have emerged, combining classical computing and QC to tackle complex scientific problems. Part of the computation is typically executed on a classical CPU, often single-threaded, while a quantum computer handles specific sub-tasks. Although QC is regarded as an extension of the frontier of HPC, few application examples in the literature take advantage of such a hybrid system. Building upon the works of Scotti~\textit{et~al.}~\cite{scotti2024clusteringaggregationalgorithmneutralatoms} and Li~\textit{et~al.}~\cite{Li2012ClusteringAA}, we propose a new hybrid HPC-QC application, and we later use it to evaluate our proposed resource management strategies.   %Many of these applications could greatly benefit from parallelizing the classical portion, enabling them to exploit the capabilities of HPC systems. 
%A representative example is the clustering aggregation hybrid algorithm developed by Scotti \textit{et al.}~\cite{scotti2024clusteringaggregationalgorithmneutralatoms} building on the original idea by Li \textit{et al.}~\cite{Li2012ClusteringAA}. 
This algorithm performs clustering aggregation to determine the optimal number of clusters based on the output of multiple clustering methods. %Its design makes it a strong candidate for hybrid HPC-QC experimentation.
%Our test case is based on the clustering aggregation method proposed by Li \textit{et al.}~\cite{Li2012ClusteringAA}, which automatically determines the optimal number of clusters and leverages multiple algorithms while demonstrating robustness to variable-quality inputs. This method was later adapted for QC by Scotti \textit{et al.}~\cite{scotti2024clusteringaggregationalgorithmneutralatoms}, who used a neutral atom quantum computer to solve a subroutine of the algorithm. 
The application exhibits several key features that make it an attractive candidate for investigating dynamic quantum-HPC resource management: 1) the classical part is highly parallelizable and could effectively capitalize on parallel execution; 2) the application scales with a variable number of classical resources; and 3) it requires quantum resources only for a limited time.
%it requires a variable number of classical resources and only needs quantum resources for a limited time. %To improve performance, we implemented a loop that iterates the same pattern, resulting in an alternating workload between classical and quantum resources.
Thus, in this work, we adapt the clustering aggregation algorithm proposed in~\cite{scotti2024clusteringaggregationalgorithmneutralatoms} to the HPC context. This adapted application serves as a use case to investigate the potential of a \emph{malleability}-based approach and a \emph{workflow}-oriented strategy for efficient resource allocation in hybrid classical-quantum environments.

\begin{figure}[]
    \centering
    \includegraphics[width=0.9\columnwidth]{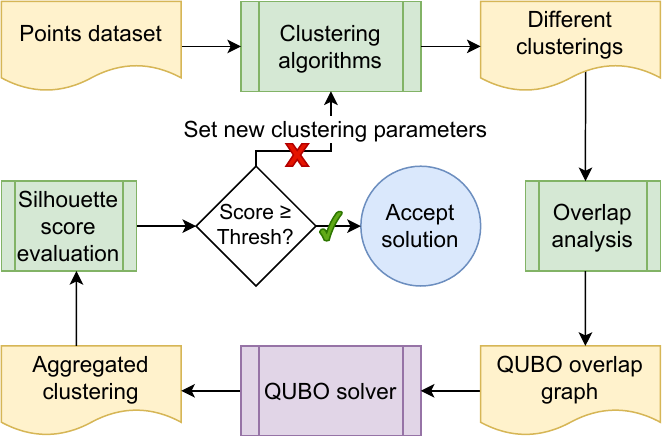}
    \caption{Data and processes of the target application. Yellow shapes are data between processes, green boxes are classical processes, and purple boxes are processes suitable to be performed by quantum resources. The application loops until the Silhouette score of a certain clustering reaches a user-defined threshold or until the maximum number of loop iterations is reached.}
    \label{fig:proposed_algorithm}
\end{figure}

Given a dataset, our C++ application, whose source code is publicly available at~\cite{clustering-mis}, begins by executing three widely used clustering algorithms: k-means~\cite{kmeans}, DBSCAN~\cite{dbscan}, and hierarchical clustering~\cite{hclust} (see Figure~\ref{fig:proposed_algorithm}). Each algorithm instance is assigned to a separate HPC node using MPI, which enables parallel execution and results in significant performance improvements over the serial baseline, assuming multiple compute nodes are available. The outputs of the three clustering algorithms are then combined to construct an undirected graph, with the constraint that a valid clustering must consist of non-overlapping clusters, as described in~\cite{scotti2024clusteringaggregationalgorithmneutralatoms}. In this formulation, each valid clustering corresponds to an independent set in the graph, and identifying a maximum independent set yields a unique and robust consensus clustering that mitigates the individual limitations of the underlying algorithms, as shown in~\cite{Li2012ClusteringAA}. Consequently, the clustering aggregation task reduces to solving a Maximum Independent Set (MIS) problem, which can be formulated as the minimization of the following function:

\begin{equation}
    f_{MIS}(x)= - \sum_{i \in \mathcal{V} } w_i x_i + \lambda \sum_{(i,j) \in \mathcal{E} } x_i x_j.
    \label{eq:wmis2}
\end{equation}
where $\lambda$ is the penalty factor, namely a corrective factor to dissuade the algorithm from choosing overlapping clusters, $w_{i}$ are weights assigned to each cluster, and $x_i \in \{0, 1 \} \forall i$.
Notably, Equation~\eqref{eq:wmis2} represents a QUBO problem.

In practice, the QUBO problem is formulated as a weighted adjacency matrix. % constructed by the rank responsible for collecting the outputs from each clustering algorithm.
The diagonal elements of this matrix contain weights $w_i$ proportional to the dimension of the corresponding cluster, so that the final solution remains balanced in terms of cluster sizes, discouraging both massive and tiny clusters. The off-diagonal entries are non-zero only for pairs of clusters that share common data points (i.e., overlapping clusters), and are assigned a value equal to the number of distinct clusters $\lambda$.

The weighted adjacency matrix is the input to the quantum portion of our application (i.e., the \emph{quantum phase}). A QUBO problem can be solved using various quantum computing architectures, including analogic quantum computers (e.g. Neutral Atoms, Quantum Annealers), and gate-based quantum computers, with the latter utilizing the Quantum Approximate Optimization Algorithm (QAOA). In this work, we solve it via Simulated Annealing (SA)~\cite{optimizationBySimulatedAnnealing}. After finding a solution to the QUBO problem, MPI rank 0 computes the silhouette score to assess the quality of the result. The silhouette value yields between $-1$ and $1$ and stands as a metric for determining the similarity of an object to its assigned cluster (cohesion) versus other clusters (separation). A high silhouette value indicates that the object is well-suited to its cluster and distinct from neighboring clusters. By evaluating the silhouette score, we choose whether to repeat the algorithms with different configurations to aim for better quality or to stop the execution. In particular, we terminate the loop when we achieve a silhouette score greater than $0.8$. To prevent an infinite loop and excessive quantum resource usage, we also set an upper bound of 10 loop iterations, after which the best result obtained up to that point is chosen.

\section{Tools for hybrid resource allocation}\label{sec:access}
%\rrnote{Maybe we can call this section tools for hybrid resource allocation?}
%\sinote{SI@All: I've added the description of DMR here, but probably it is not the best place to be.}

The structure of the application flow defines two segments that must run on different hardware. From an HPC center perspective, providing access to multiple applications with this pattern is non-trivial, as quantum resources are currently scarce. The difficulty behind resource allocation comes from the imbalance between the requests for quantum resources and their availability, which inevitably introduces a bottleneck in the system. To ease the bottleneck, we must ensure that the users access scarce resources properly to avoid reserving them when unneeded, thus blocking the execution for all the others.
In general, requesting both classical and quantum resources in a SLURM-based fashion at the beginning of the execution and for its entire duration is inefficient, as the application uses only one type of resource at any time.
%Additionally, we are blocking scarce resources without actually using them all the time, worsening the bottleneck for the entire cluster.
On the other hand, reserving only classical resources and relying on a Representational State Transfer (REST) Application Programming Interface (API) to access the quantum node, as is commonly done in today's cloud-based quantum systems, is not efficient:
%offload systems like the ones used in the QC field via cloud nowadays is still inefficient:
while we wait to schedule the quantum offloaded portion, we are holding classical resources without using them. %This latter aspect impacts less than the former; nonetheless, classical resources are not infinite, and other users (not interested in quantum applications) may use them too.

The difficulty in maintaining efficiency comes from the discrepancy between the changing resource requirements of hybrid applications and the staticity of resource allocation. This problem is not new and already features some theorized solutions~\cite{Simone,beck2024quantum,mantha_2024,Elsharkawy_2024,osti_2006942,saurabh_2023}. Out of all the proposed approaches, we think that a \emph{workflow} implementation of the execution flow and enabling dynamicity through \emph{malleability} can yield significant benefits regarding efficient resource allocations. In the following subsections, we describe these two approaches, presenting some tools we leverage to implement those functionalities.

\subsection{Workflow}

%\gbnote{Here a paragraph about workflow in general @Paolo. 
%Then we focus on Streamflow  @Iacopo}
%\rrnote{
%1) Workflow in general (HPC)
%2) Workflow in HPC-QC~\cite{CRANGANORE2024346}
%3) Workflow with StreamFlow
%4) Using Streamflow for our app}

A workflow approach divides an application into tasks and executes them according to the dependencies between them~\cite{Suter:2026}. Describing complex, large-scale applications as workflows dramatically simplifies their transition from classical execution to hybrid HPC-QC settings. Various workflow engines have been developed to orchestrate the execution of applications across heterogeneous computing environments (e.g., Nextflow~\cite{di2017nextflow}, PyCOMPSs~\cite{pycompss}, and Rigoletto~\cite{DeMaio_2024}).

The StreamFlow WMS~\cite{Colonnelli:2021} has been explicitly designed with \emph{hybrid workflows} in mind. A hybrid workflow can be expressed as $(W,$$L,$$\mathcal{M})$, where $W$$=$$(S,$$P,$$\mathcal{D})$ is a directed bipartite graph representing a standard workflow (with steps $S$, ports $P$, and dependency links $\mathcal{D}$), $L$ is the set of available execution locations, and $\mathcal{M}$$\subset$$(S$$\times$$L)$ is a mapping function \cite{Colonnelli:2024}. This abstraction is general enough to encode workflow models specifically targeting classical-quantum applications. For example, in~\cite{CRANGANORE2024346}, workflow designers are required to identify a set $Q$$\subseteq$$S$ of \emph{quantum candidates}, i.e., steps that support a quantum implementation that is \emph{functionally equivalent} to the classical one~\cite{CRANGANORE2024346}. At runtime, the WMS chooses between the classical and quantum implementation, depending on the availability of quantum resources. This technique can be encoded as a hybrid workflow by identifying a subset of \emph{quantum locations} $L_{Q}$$\subseteq$$L$ and defining a \emph{quantum mapping} function $\mathcal{M}$$\subseteq$$(Q$$\times$$L_Q)$. The principal advantage of this more general approach is that it allows a seamless transition from specialized HPC-QC orchestrators to X-QC settings, including cloud-QC and HPC-cloud-QC. 

%the StreamFlow WMS has been explicitly designed around the concept of hybrid workflows~\cite{Colonnelli:2021, Colonnelli:2023, Colonnelli:2024}.
Implementation-wise, StreamFlow augments the Common Workflow Language (CWL)~\cite{CWL:2022} open standard with a topology of deployment locations and a \texttt{Loop} extension to model iterative workflows~\cite{ml4astro:22}. The orchestration plane leverages pluggable connectors to support several execution environments, from HPC queue managers to cloud infrastructures. Developing a new plugin to support a given quantum device is just a matter of extending the StreamFlow \texttt{Connector} base class to integrate the device's APIs, whether web-based REST APIs or SLURM-managed queues.

These features make StreamFlow a natural fit for implementing our HPC-QC application as a hybrid workflow. The resulting model is a three-step pipeline that computes and combines the clusterings (step $s_{C}$), aggregates them by solving a QUBO problem (step $s_{Q}$), and evaluates the resulting silhouette score (step $s_{S}$), s.t. $S$$=$$\{s_{C},$$s_{Q},$$s_{S}\}$. Each stage is implemented as a CWL \texttt{CommandLineTool} object, and the coordination logic is encoded as an iterative CWL \texttt{workflow} that repeats the whole process until the silhouette score reaches a configurable threshold. The execution environment is composed of three classical locations (HPC nodes) and one quantum location, s.t. $L$$=$$\{l^{c}_{1},$$l^{c}_{2},$$l^{c}_{3},$$l^{q}_{1}\}$. The resulting mapping function is $\mathcal{M}(s_{C})$$=$$\{l^{c}_{1},$$l^{c}_{2},$$l^{c}_{3}\}$, $\mathcal{M}(s_{Q})$$=$$l^{q}_{1}$, and $\mathcal{M}(s_{S})$$=$$l$$\in$$\{l^{c}_{1},$$l^{c}_{2},$$l^{c}_{3}\}$.
Note that this example can also be modeled in the Cranganore \textit{et al.}~\cite{CRANGANORE2024346} framework by defining $Q$$=$$\{s_{Q}\}$ and $L_{Q}$$=$$\{l^{q}_{1}\}$.
\subsection{Malleability}

Malleability refers to the ability of a parallel application to dynamically adjust its~\cite{feitelson_packing_1996} resource allocation, such as the number of computing nodes or processes, during runtime. This flexibility enables jobs to grow or shrink in response to changing system conditions, improving overall resource utilization and reducing wait times~\cite{iserte_towards_2024}. Malleability is an active area of research in traditional HPC. Among the several tools available to aid the implementation of malleability~\cite{aliaga_survey_2022}, 
%~\cite{huang2003adaptive,martin2015enhancing,pickartz2019virtualization,huber2024designprinciplesdynamicresource}. 
the DMR framework~\cite{iserte_dmrlib_2020} provides a high-level API designed to incorporate malleability into HPC applications seamlessly.
DMR is a well-known tool in the community~\cite{tarraf_malleability_2024}, has been thoroughly evaluated across a variety of use cases~\cite{iserte_study_2020, iserte_dynamic_2018, iserte_dmr_2018}, and has also been extended with novel functionalities to support broader capabilities~\cite{huber_bridging_2025, iserte_resource_2025}.
DMR establishes communication between the Parallel Distributed Runtime (PDR), the Resource Management System (RMS), and the Parallel Performance Monitor (PPeM) to enable transparent process and resource management.

%\begin{figure}[t]
%    \centering
%    \includegraphics[width=0.9\linewidth]{Fig/dmr.png}
%    \caption{DMR scheme and configuration.}
%    \label{fig:dmr}
%\end{figure}

%Figure~\ref{fig:dmr} depicts how DMR connects the application with the rest of subsystems leveraged in this work.
At its core, DMR enables passive dynamic resource management through the following process:

\begin{itemize}
\item During execution, DMR periodically checks when jobs are ready for reconfiguration at predefined synchronization points within the application.
\item DMR determines whether a reconfiguration is necessary, considering available resources and performance metrics in collaboration with the RMS and the PPeM.
\item DMR orchestrates the updates in resource allocations (interacting with the RMS) and the number of processes (through the PDR).
\item Finally, the execution resumes seamlessly from the reconfiguration point with the new job shape.
\end{itemize}

%DMR has established itself as a robust and adaptable malleability tool~%its effectiveness has been validated %through numerous published success stories, among others those in~\cite{%iserte_dynamic_2018,iserte_study_2020,
%\cite{iserte_towards_2024}, % , and several ongoing developments. 
%These contributions continue to highlight its ability to 
%enhancing efficiency and flexibility in HPC applications across diverse computing environments.
%DMR has established itself as a robust and adaptable malleability tool~\cite{iserte_towards_2024}, enhancing efficiency and flexibility in HPC applications across diverse computing environments.

Listing~\ref{code:dmr} shows the structure of the implemented HPC-QC malleable code described in Section~\ref{sec:hybrid}. First, the DMR environment is initiated and provided with the program-specific restart function, used by new processes after reconfigurations (line~3). 
The main loop (line~4) has two stages: classical and quantum. At the beginning of each iteration, if the execution has to run the classical stage, a reconfiguration can be scheduled (line~7). 
After this stage, only the root MPI process will be kept active, relinquishing the remaining resources (line~10). The reconfiguration requires an application-specific checkpoint function to save the current state, which comprises the current loop iteration number and a flag that specifies whether we can proceed to the quantum stage. This state is then restored by the restart function.
Eventually, the execution continues in the quantum stage (line~13) without wasting classical resources.

%Adding malleability properties in the application described in Section~\ref{sec:hybrid} quite natural. Listing~\ref{code:dmr} exemplifies the structure of the HPC-QC malleable code implemented in this work.
%Initially, the malleability environment has to be defined (line~3).
%Afterwards, we initialize the DMR library and provide a program-specific restart function that new processes execute after a reconfiguration to restore their state to the point where they began reconfiguring.
%The main loop (line~4) is composed of two stages: classical and quantum.
%At the start of each iteration, if the classical stage is to execute, a reconfiguration may be launched (line~7) if additional resources are needed and have been secured.
%At the end of the classical stage, the job releases the maximum amount of classical resources (line~10) in order to execute the quantum stage (line~13), leaving a single classical process alive to monitor progress. Note that to reconfigure, we have to provide DMR with an application-specific checkpoint function to save the current status of the execution, which is later read back by the aforementioned restart function.

\begin{lstlisting}[float, caption=Scheme of how malleability is adopted in the user code., label=code:dmr, captionpos=b]
void main() {
	MPI_Init();
	DMR_INIT (restart(it0, quantum));
	for (auto it = it0; it < ITERS; it++) {
		if (!quantum) {
			while (!resources && !timeout)
				DMR_EXPAND (checkpoint(it, quantum));
		  //Execute Classical Computation
		  quantum = true;
		  DMR_MINIMIZE (checkpoint(it, quantum));
		}
		if (rank_id == 0) 
			//Offload Quantum Computation
		quantum = false	
	}
	MPI_Finalize();
}
\end{lstlisting}

\section{Experimental Campaign}\label{sec:exp}
We conducted a series of experiments by executing our hybrid clustering application under different configurations to evaluate the impact of the solutions described in Section~\ref{sec:access}. In the following paragraphs, we present the experimental setup and overview of the campaign, analyzing its results.

\noindent\textbf{Experimental setup}. We set up a dedicated cluster to test our approaches, using SLURM version 23.02.7. The cluster represents a plausible HPC-QC integration scenario at scale, comprising two partitions: a log-in and a master node. The first partition, named \emph{compute}, consists of three CPU-only nodes. Each compute node contains two AMD EPYC 7543 CPUs and $256 \, \mathrm{GB}$ of DDR4 memory. The second partition, named \emph{quantum}, acts as a quantum emulator. This node contains two AMD EPYC 7282 CPUs, $512 \, \mathrm{GB}$ of DDR4 memory. However, due to the memory requirements of the clustering algorithms, we launch our application requesting one or more classical nodes and one task per node.
%, and two NVIDIA Tesla V100S GPUs with $32 \, \mathrm{GB}$ of HBM2 memory each.
The log-in node and the master node are virtualized x86\_64 machines with $8 \, \mathrm{GB}$ of memory each. Every machine in the described cluster runs on Red Hat Enterprise Linux 9.4, using kernel version 5.14. The application source code is compiled with GNU Compiler Collection version 11.4.1. The distribution of classical workloads across the compute partition is achieved with OpenMPI version 5.0.6. Submitting quantum jobs is carried out with HyperQueue~\cite{hyperqueue} version 0.21.1. In particular, HyperQueue allocates a node from the SLURM quantum partition and launches the QUBO solver on it. StreamFlow version 0.2.0.dev12 is used to validate the workflow approach.

%\noindent\textbf{Quantum emulation}.
\noindent\textbf{Towards quantum}. In our pipeline, the quantum phase is formulated as a QUBO problem. We currently solve it using Simulated Annealing on a classical node, which mimics the behavior of a Quantum Annealing algorithm. Since the problem core formulation remains unchanged, substituting the classical node, initially a placeholder, with an actual quantum device requires no modifications to the proposed configurations or methodologies. With real quantum hardware, such as a superconducting quantum processor, a quantum annealer, a neutral atom device, or a photonic quantum computer, the QUBO problem can be tackled using inherently quantum approaches.
%The quantum phase reduces (in our pipeline) to solve a QUBO problem, and we perform a Simulated Annealing on the classical node to mimic the Quantum Annealing algorithm. Replacing the classical node that we used as placeholder of a quantum machine with an actual one, we can mantain the proposed cluster configuration. Through to a real quantum hardware we could solve the QUBO with a quantum approach. Many quantum machines superconducting quantum machine, or a quantum annealer, or a neutral atom device, or a photonic computer allow to solve a QUBO. %Since we cannot access real quantum hardware in our testing environment, we emulate a quantum computer by performing Simulated Annealing (SA)~\cite{optimizationBySimulatedAnnealing} on a classical node,
%SA is a probabilistic method for finding the minimum of a given input function, and it 
%behaving like quantum annealing in our context. % More precisely, they both try to find the minimum of a given input function.
%Since our focus is on the scheduling of resources and not on the optimization quality, we can emulate quantum annealing by using SA and by adding artificial delays when appropriate to account for the expected runtime of quantum annealing on real quantum hardware. 
%As an alternative to the SA algorithm, we could have also employed the QAOA using CUDA-Q to solve the minimization problem. Additionally, When available, utilizing a real quantum device is another viable option. 
It is essential to note that different quantum technologies exhibit distinct characteristics and execution times. For instance, a neutral atom device would typically require minutes to complete a job (also considering register preparation), whereas a superconducting device operates on a much faster timescale, in the order of seconds. % In future work, we aim to explore the feasibility of incorporating these real quantum approaches, if possible.
To simulate the performance of a neutral atom device, we introduce an artificial delay of some minutes in those runs, %the present study,
allowing us to analyze and understand the potential benefits and limitations of dynamic resource allocation in such systems.
The duration of these artificial delays is determined based on the quantum machines of interest, considering the typical preparation and execution times of their average quantum job.

\subsection{Experimental Results}
We execute a set of comparative experiments to evaluate the behavior of a system running hybrid HPC-QC workloads. We begin by analyzing a configuration that adopts a traditional resource management strategy, allocating classical resources for the entire duration of the hybrid job and offloading the quantum subroutine to the emulated QPU using HyperQueue. This configuration is referred to as the \emph{baseline} throughout the rest of the paper, as it resembles the current offloading scheme diffused with cloud-based quantum machines. We then compare the baseline results with those of the two alternative approaches for hybrid resource management discussed in Section~\ref{sec:access}: \emph{workflow} and \emph{malleability}. As for the baseline, the malleability offloads the quantum task via HyperQueue, whereas the workflow approach independently allocates a node from the quantum partition only when required. For all three configurations, we evaluate and compare the following metrics:
%In particular, we first analyze a solution, which employs a traditional resource management strategy to allocate classical resources for the entire duration of the hybrid job and HyperQueue to offload the quantum subroutine to the emulated QPU. This is referred to as \emph{baseline} in the rest of the text. Then we compare these results with the \emph{workflow} and \emph{malleability}-based approaches for managing hybrid resource allocation. Both the baseline and malleability approaches offload quantum tasks using HyperQueue, while the workflow approach independently allocates a node from the quantum partition when required. In particular, we examine the following metrics across the three scenarios discussed above:
\begin{itemize}
    \item Wall time: the time employed to execute the whole simulation, considering also the time spent by SLURM to initialize and finalize the job(s);
    \item Classical resource usage in terms of node-seconds: the sum of the product between the number of classical nodes used in an interval and the length (in seconds) of the interval (we disregard here the quantum resources because their usage is constant across the three scenarios);
\end{itemize}

%To explore the impact of \emph{malleability} on hybrid HPC-QC workloads, we compare this approach with what we called \emph{baseline}. This consists in the usage of a traditional resource management method, which \textcolor{blue}{co-allocate classical and quantum resources for the duration of the whole hybrid job}. Additionally, we compare our malleability-based solution with another possible solution for managing hybrid resource allocation: \emph{workflow}. The baseline and the malleability approaches offload quantum jobs using HyperQueue, whereas the workflow approach allocates a node from the quantum partition independently.

%, and \emph{malleability}. We evaluate their advantages with respect to the so-called \emph{baseline} approach that implements a static allocation of the classical part. The baseline and the malleability approaches offload quantum jobs using HyperQueue, whereas the workflow approach allocates a node from the quantum partition independently.

The dataset used in the clustering application consists of 80,000 2D points generated via \texttt{make\_blobs} from scikit-learn\footnote{\url{https://scikit-learn.org/stable/modules/generated/sklearn.datasets.make_blobs.html}}. The application code is written to permit reproducible results. In particular, given the described dataset, each workload run ends after the fourth loop iteration, as the combined clusterings achieve a silhouette score over 0.8.
We profile two versions of the application to better understand how different quantum technologies might suit specific scheduling approaches. A first version executes quantum jobs in a fraction of a second, thus akin to superconducting QPUs. A second version adds an artificial delay of two minutes during the quantum job, trying to mimic a generic neutral atom QPU.% The delay does not intend to replicate the execution time of a specific neutral atom QPU, and it is only chosen to investigate the impact of quantum jobs of the order of minutes.

We start from the executions with no resource contention, i.e. with no other jobs in the cluster queue, and with the two-minute-long quantum jobs, i.e. reproducing the behavior of a neutral atoms machine. We average the metrics from five runs for each strategy.
Table~\ref{table:medium_jobs_combined} shows the result of the first experiment. The baseline approach is the fastest one, but it is less efficient regarding resource usage. The workflow approach performs poorly in terms of wall time since it asks SLURM for resources at every step, and the overhead of the WMS slows it down. Conversely, it is the best regarding resource usage with minimal node-second consumption. The malleability approach acts as a compromise between the other two.
%\textcolor{blue}{Check the following paragraph carefully} On one hand, resource utilization is slightly higher than when the application is executed using the workflow approach ($\approx +29 \%$), but significantly lower compared to the classical allocation strategy ($\approx -46 \%$). On the other hand, as expected, the application runs slightly slower than with the baseline approach ($\approx +1 \%$ in time-to-solution), but completes faster than the workflow approach ($\approx -3 \%$). 
In the absence of resource contention, both malleability and workflow approaches primarily conserve valuable computational resources with a negligible impact on time-to-solution.

%\begin{table}[]
%\caption{Executions with medium-length quantum jobs (2 minutes) and no resource contention.}
%\vspace{-5pt}
%\label{table:single_workload}
%\begin{tabular}{l|l|l|}
%\cline{2-3}
%                                   & Wall time [seconds]      & Resource usage [node-seconds] \\ \hline
%\multicolumn{1}{|l|}{Baseline}     & 1019.58 $\pm$ 0.85                   & 3058.74 $\pm$  2.56                                  \\ \hline
%\multicolumn{1}{|l|}{Workflow}     & 1057.80 $\pm$ 6.02 & 1161.20 $\pm$ 6.94                       \\ \hline
%\multicolumn{1}{|l|}{Malleability} & 1029.06 $\pm$ 1.54 & 1647.75 $\pm$ 1.54                  \\ \hline
%\end{tabular}
%\end{table}

\begin{figure}[t]
    \centering
    \includegraphics[width=0.9\columnwidth]{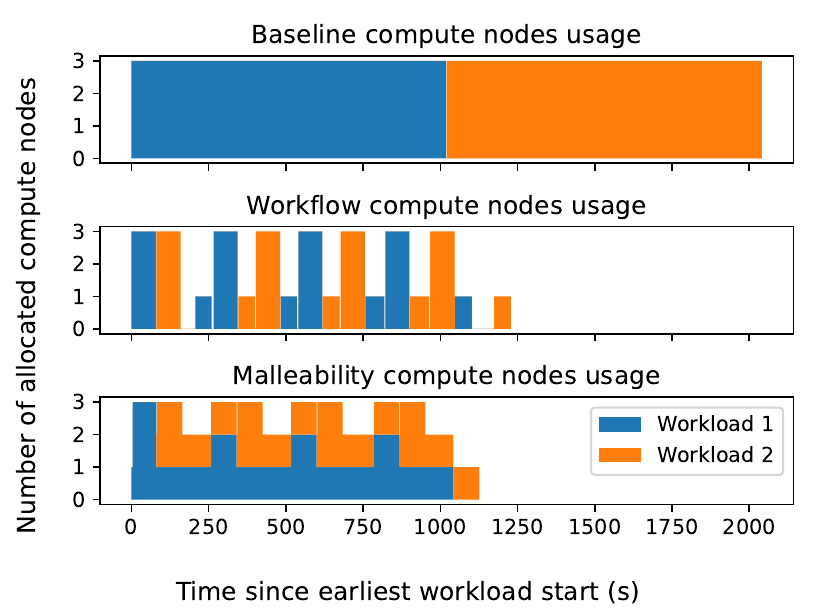}
    \caption{Timeline of cumulative usage of classical nodes on our SLURM compute partition when launching two concurrent workloads (in blue and orange, respectively) with two minutes long quantum jobs. %Can be commented if necessary
    Each subplot refers to a different scheduling approach, i.e. baseline, workflow, and malleability.} %The overlapping areas in the picture represent the sum of the resource usage of the two jobs.}
    %\vspace{-12pt}
    \label{fig:comparison_dual_workloads_2min_sleeps}
\end{figure}

For our second experiment, we run two concurrent workloads using all the approaches, again under a queue empty from other submissions and by emulating two-minutes-long quantum jobs. %\textcolor{blue}{As usual,} we force each process to request exclusively one node, effectively posing a limit to the concurrent classical processes to three. This eases the testing of jobs that compete for classical resources, since two jobs are enough to study contention.
Table~\ref{table:medium_jobs_combined} contains the results of this second experiment, averaged over five runs each. Note that the wall time here refers to the elapsed time between the beginning of the first starting workload and the completion of the latest ending workload, thus considering two complete end-to-end simulations. The baseline approach is, in this case, the worst-performing one. The other approaches can interleave their execution, finishing earlier and using fewer resources, as shown in Figure~\ref{fig:comparison_dual_workloads_2min_sleeps}.
%\textcolor{blue}{In particular, the use of malleability in this case results in completion time of both simulations $\approx 44\%$ shorter than the baseline and $\approx 8\%$ shorter than the workflow approach. It also reduces node-second usage by $\approx 54\%$ compared to the baseline, although it consumes around $\approx 17\%$  more resources than the workflow. 
The difference between malleability and workflow results resides in the need for the former to have at least one MPI process to remain active at all times, even during the quantum phase when computations are offloaded to the QPU. While this may seem like an overhead, it offers a clear advantage: when the code returns from the quantum phase, execution can resume immediately, allowing the simulation to proceed even if not all originally requested resources are available. %A graphical comparison of this experiment is reported in Figure~\ref{fig:comparison_dual_workloads_2min_sleeps}, which shows the differences in resource usage. Workflow and malleability approaches can efficiently use classical resources by interleaving classical parts execution, while the baseline approach cannot do so.

We then execute the experiments above without artificial delays in the quantum phase. The results of these new experiments can be seen in Table~\ref{table:single_workload_no_sleep}. For baseline and workflow, the wall time and resource usage of the dual concurrent execution are almost double the values from the single execution. This means the classical resources are highly contested, with little space to optimize scheduling. The malleability solution, on the other hand, completes both simulations in a time comparable to that of a single execution ($648.61 \, \mathrm{s}$ vs $549.60 \, \mathrm{s}$), demonstrating its ability to manage and adapt to varying resource demands dynamically. However, this result comes from the intrinsic properties of the algorithms executed during the classical phase (with k-means taking significantly less time than the other two), so they do not represent a trend. In general, scenarios involving short quantum jobs and no resource contention should show limited resource savings when using either workflow or malleability approaches. % are less significant compared to the case with medium-length quantum jobs. 

\begin{table}[]
\centering
\caption{Executions with $2$ minutes long quantum jobs.}
\label{table:medium_jobs_combined}
\begin{tabular}{|l|c|c|c|}
\hline
Execution Type & Mode & \begin{tabular}[c]{@{}c@{}}Wall time\\ {[}seconds{]}\end{tabular} & \begin{tabular}[c]{@{}c@{}}Resource usage\\ {[}node-seconds{]}\end{tabular} \\ \hline
Baseline       & Single & 1019.58 $\pm$ 0.85 & 3058.74 $\pm$ 2.56 \\
Workflow       & Single & 1057.80 $\pm$ 6.02 & 1161.20 $\pm$ 6.94 \\
Malleability   & Single & 1029.06 $\pm$ 1.54 & 1647.75 $\pm$ 1.54 \\
\hline
Baseline       & Double & 2038.43 $\pm$ 0.96 & 6115.30 $\pm$ 2.89 \\
Workflow       & Double & 1226.00 $\pm$ 1.58 & 2324.00 $\pm$ 3.39 \\
Malleability   & Double & 1127.65 $\pm$ 1.18 & 2817.73 $\pm$ 1.27 \\ \hline
\end{tabular}
%\vspace{-10pt}
\end{table}

\begin{table}[]
\centering
\caption{Executions with short ($<1$ second) quantum jobs.}
%\vspace{-5pt}
\label{table:single_workload_no_sleep}
\begin{tabular}{|l|c|c|c|}
\hline
Execution Type & Mode & \begin{tabular}[c]{@{}c@{}}Wall time\\ {[}seconds{]}\end{tabular} & \begin{tabular}[c]{@{}c@{}}Resource usage\\ {[}node-seconds{]}\end{tabular} \\ \hline
Baseline       & Single & 539.44 $\pm$ 0.53 & 1618.33 $\pm$ 1.60 \\
Workflow       & Single & 569.00 $\pm$ 3.94 & 1148.00 $\pm$ 1.87 \\
Malleability   & Single & 549.60 $\pm$ 1.86 & 1168.29 $\pm$ 1.81 \\
\hline
Baseline       & Double & 1076.98 $\pm$ 1.79 & 3230.95 $\pm$ 5.37 \\
Workflow       & Double & 1089.00 $\pm$ 1.00 & 2324.00 $\pm$ 4.24 \\
Malleability   & Double & 648.61 $\pm$ 2.08 & 1622.63 $\pm$ 1.05 \\ \hline
\end{tabular}
\end{table}

%\begin{table}[b]
%\caption{Executions with short quantum jobs and 2 concurrent workloads.}
%\label{table:double_workload_no_sleep}
%\begin{tabular}{l|l|l|}
%\cline{2-3}
%                                   & Wall time [seconds]      & Resource usage [node-seconds] \\ \hline
%\multicolumn{1}{|l|}{Baseline}     & 1076.98 $\pm$ 1.79 & 3230.95 $\pm$ 5.37 \\ \hline
%\multicolumn{1}{|l|}{Workflow}     & 1089.00 $\pm$ 1.00 & 2324.00 $\pm$ 4.24 \\ \hline
%\multicolumn{1}{|l|}{Malleability} & 648.61 $\pm$ 2.08 & 1622.63 $\pm$ 1.05 \\ \hline
%\end{tabular}
%\end{table}

\subsection{Discussion}
Experimental data suggests that malleability and workflow are well-suited for hybrid classical-quantum workloads with long-running quantum phases.
%As such, they could be advantageously used with neutral atom quantum computers. 
%\textcolor{blue}{This must be checked with the number. Currently, I only soften what was written here. However, in the case of superconducting quantum computers the benefits are perhaps less evident}. 
The specific choice between a workflow or a malleability approach depends on what a user values the most: if resource usage is crucial, workflows allow for using resources very efficiently; when time to solution is more critical, a malleability approach could be more appropriate, especially in HPC clusters with long queue times. 
Thus, even if somewhat anticipated, these results highlight the relevance of this work as part of a still limited but growing set of practical efforts addressing hybrid HPC-QC scheduling (e.g., \cite{Wille_2024,esposito2025slurmheterogeneousjobshybrid, tejedor2025distributedquantumcircuitcutting}).
Furthermore, while our current metrics do not account for queue times, incorporating this factor would further highlight the advantage of malleable jobs in reducing queuing delays, enabling users to obtain results more quickly.% \textcolor{blue}{To further increase the generality and impact of our findings, these strategies should be validated on a broader set of more complex hybrid HPC-QC workloads and possibly on real quantum hardware. All these tasks are left for future exploration.}

We note that neither of the proposed approaches is a drop-in replacement inside an application with fully static resource allocation. Workflows require the programmer to create modular applications instead of monolithic ones. This would likely improve the quality of the code, but the user would need to understand a dedicated workflow language to glue the application parts together. Conversely, we argue that malleability approaches have low entry barriers since the offered interface is user-friendly for programmers, but they have to manage a code base with more complex state management and execution with variable resources. 

\section{Conclusion}\label{sec:conclusion}

%\rrnote{TODO tomorrow. Structure in 3 paragraphs, each with a couple short periods. First about experimental results, then about our contributions, then about problem as a whole. Mention future work at the end, but briefly.}
%\gbnote{the novelty of the work relies on the malleability approach, we should underline this}

%In this work, we present an analysis of the resource allocation problem present in the current HPC-QC field, addressing it through a novel malleability solution and a workflow approach. Using malleability properties in the HPC-QC field has never been theorized before, but has the potential to yield remarkable benefits. We analyze a specimen use-case of HPC-QC application and show the effectiveness of malleability and workflow approaches compared to a statically-allocating baseline.
In this work, we analyze the resource allocation challenges in the emerging HPC-QC landscape and propose a novel solution based on malleability alongside a workflow-based approach. To our knowledge, malleability has not previously been explored in the context of HPC-QC, yet it holds significant potential for improving efficiency. We demonstrate the benefits of malleability and workflow strategies through a representative HPC-QC application, showing clear advantages over a traditional, statically allocated baseline.
Our experiments show that, under resource contention, malleability reduces execution time by approximately $\approx$~$44\%$ compared to the baseline and by $\approx$~$8\%$ compared to the workflow approach. Regarding resource consumption, malleability achieves a $\approx$~$54\%$ reduction in node-seconds compared to the baseline while incurring a $\approx 17.5\%$ overhead compared to the workflow approach.

Managing a current HPC-QC cluster is not trivial, as the classical and quantum partition differences go beyond the simple programming model. As the maturity of the quantum software stack grows, the scientific community is also considering the issues coming from resource allocation. The present work contributes to this effort by leveraging the current state of the art in supercomputing systems to bridge the gap until more advanced, integrated frameworks become available. While we do not claim to offer a one-size-fits-all solution, we propose dynamic resource allocation strategies, both demonstrating performance improvements over the baseline.
%Although we did not design a one-fits-all solution, we proposed two possible solutions for dynamic resource allocation, improving the performances with respect to the baseline. %Additionally, the heterogeneity of the quantum machines spectrum implies that it is difficult to design a one-fits-all solution for resource allocation. 
%
%, or we would highly compromise the usability of HPC-QC machines. 

%Our experiments demonstrate that, with resource contention, malleability leads to execution times $\approx 44\%$ and $\approx 8\%$ shorter than the baseline and the workflow, respectively. In terms of resource usage, malleability allows for saving $\approx 54\%$ node-seconds compared to the baseline, although it consumes around $\approx 17\%$  more resources than the workflow. 

Future work in this area includes extending this work by considering a larger-scale execution environment with a complex resource contention scenario, possibly addressing an actual quantum machine. % \textcolor{blue}{and more complex hybrid HPC-quantum workloads}.

%Future work:
%\begin{itemize}
%    \item Perform all the previous tests while the cluster is not empty (consider random jobs or other techniques...). When considering a non-empty cluster we should consider that the experiment has to be reproducible so random jobs might be not the best idea. 
%    \item  consider also the waiting time as metric (it should be an information we could retrive from SLURM)
%    \item novelty f our approach: employing malleability that should show its advantages when...
%    \item try the approach with real quantum machines (as said previous section (quantum emulation)
    
%\end{itemize}
%\gbnote{Short papers are 4-6 pages (including figures and tables and appendices), plus one additional page of references.}

%\section*{Acknowledgment}

%The preferred spelling of the word ``acknowledgment'' in America is without 
%an ``e'' after the ``g''. Avoid the stilted expression ``one of us (R. B. 
%G.) thanks $\ldots$''. Instead, try ``R. B. G. thanks$\ldots$''. Put sponsor 
%acknowledgments in the unnumbered footnote on the first page.

\bibliographystyle{IEEEtran}
\bibliography{bibliography}
\end{document}